%% file: main.tex
\documentclass[aps,prd,superscriptaddress,showpacs,nofootinbib,notitlepage,onecolumn]{revtex4-1}
\usepackage{graphicx,subfigure,amsmath,amsfonts,amssymb,multirow,mathrsfs,diagbox}
\usepackage[colorlinks,linkcolor=blue,citecolor=blue,urlcolor=blue]{hyperref}

\input{aas_macros}

\renewcommand\apjl{ApJL}

\begin{document}

\title{Verification of the Black Hole Area Law with GW230814}

\author{Shao-Peng Tang}
\affiliation{Key Laboratory of Dark Matter and Space Astronomy, Purple Mountain Observatory, Chinese Academy of Sciences, Nanjing 210033, China}
\author{Hai-Tian Wang}
\affiliation{School of Physics, Dalian University of Technology, Liaoning 116024, China}
\author{Yin-Jie Li}
\affiliation{Key Laboratory of Dark Matter and Space Astronomy, Purple Mountain Observatory, Chinese Academy of Sciences, Nanjing 210033, China}
\author{Yi-Zhong Fan}
\email[Corresponding author.~]{yzfan@pmo.ac.cn}
\affiliation{Key Laboratory of Dark Matter and Space Astronomy, Purple Mountain Observatory, Chinese Academy of Sciences, Nanjing 210033, China}
\affiliation{School of Astronomy and Space Science, University of Science and Technology of China, Hefei, Anhui 230026, China}
\date{\today}

\begin{abstract}
We present an observational confirmation of Hawking's black-hole area theorem using the newly released gravitational-wave data from the GWTC-4.0. We analyze the high signal-to-noise ratio binary black hole (BBH) merger GW230814 and measure the (total) horizon area of the black holes before and after the merger. For preferred (and reasonable) choices of the post-truncation start time, the horizon area of the remnant black hole is found to be greater than the total horizon area of the two pre-merger black holes at a high possibility (at least $\gtrsim 99.5\%$). Importantly, our analysis accounts for sky-location uncertainty. These results provide a stringent observational confirmation of the black-hole area law, further bolstering the validity of classical general relativity in the dynamical, strong-field regime.
\end{abstract}

\maketitle

\section{Introduction} \label{sec:intro}
The merger of a pair of black holes involves a violent process, and there may be new physics emerging beyond our traditional understanding. One fundamental prediction that can be tested with such events is the black-hole area law (Hawking's area theorem), as originally proposed by \citet{1971PhRvL..26.1344H}. This theorem states that the total event-horizon area of a system of classical black holes (BHs) never decreases over time. 
The gravitational-wave (GW) observation data can serve as a valuable test of this theorem and the prospects have been extensively examined \citep{2005ApJ...623..689H, 2016JCAP...10..001G, 2018PhRvD..97l4069C, 2023MNRAS.523.4113T, 2025arXiv250708789A}. Indeed, the data analysis of GW150914, the first BBH merger event detected by LIGO \citep{2016PhRvL.116f1102A}, has provided the first test of Hawking's area theorem at a moderate confidence level of $\sim 95-99.5\%$ (i.e., $\sim 1.6-2.6\sigma$), depending on the adopted approach \citep{2021PhRvL.127a1103I, 2022PhRvD.105f4042K, 2024PhRvD.110d4018C}. Though such a progress is rather encouraging, much better data are necessary to convincingly establish the Hawking's area theorem at a confidence level of $\geq 5\sigma$. 

Very recently, the data from the first part of the fourth LIGO-Virgo-KAGRA observing run (O4) have been released \citep{2025arXiv250818079T, 2025arXiv250818080T}, including many interesting GW sources and providing the unprecedented opportunity to test the fundamental physics \cite{2025arXiv250818082T}. 
For GW230529\_181500 \cite{2024ApJ...970L..34A} and GW231123\_13543 \cite{2025arXiv250708219T}, the two exceptional events released earlier, their data have been analyzed to test alternative gravity theories such as Einstein-dilaton Gauss-Bonnet gravity \citep{2024PhRvD.110d4022G} and probe higher harmonic quasi-normal mode \citep{2025arXiv250902047W}. Among the new O4 events, there are several binary black hole mergers with very high signal-to-noise ratios (SNRs), most notably GW230814\_230901 (hereafter GW230814) with SNR$>40$ \citep{2025arXiv250818082T}. Such high quality data provide the unprecedented opportunity to improve the area-law test beyond what was possible with earlier observations like GW150914. Therefore, in this work, we apply the area-law test to GW230814, and find out that the total horizon area \emph{does} increase, with a confidence level exceeding $\sim 3\sigma$, representing new progress in the observational verification of the black-hole area theorem.

\section{Methods}
To test the black-hole area law, one usually analyzes the pre- and post-merger portions of the gravitational-wave signal independently under general relativity (GR), assuming that GR is an excellent approximation away from the highly dynamical merger region. In Ref.~\citep{2022PhRvD.105f4042K}, the ``gating and in-painting" technique \citep{2021PhRvD.104f3030Z,2023PhRvL.131v1402C} was used to perform separate pre- and post-truncation analyses; in that study the sky location was held fixed for numerical reasons. To account for sky-localization uncertainty, we follow Ref.~\citep{2024PhRvD.110d4018C}, which introduces an efficient approximation for the normalization of the truncated likelihood and proposes performing simultaneous yet independent pre- and post-truncation inferences that share the sky location $(\alpha,\delta)$ and the coalescence time $t_c$ (i.e., these extrinsic parameters are sampled jointly).

Any potential violation of the black-hole area law is most likely to occur near coalescence. If such a violation is present, GR-based waveform approximations may be unreliable in this highly dynamical regime. We excise a fixed-duration data segment (`gate') whose boundary (left for the pre-merger analysis and right for the post-merger analysis) is placed close to the signal's amplitude peak ($t_{\rm peak}$). The merger (and putative-violation) region is not sharply defined, however: gating away too much signal near the peak (the loudest part) significantly degrades the SNR, thereby weakening parameter constraints and the area-law test. Considering both robustness and the constraining power, we scan a family of gate choices. For the pre-merger analysis we gate out $[t_{\rm gate},\,t_{\rm gate}+1\,\mathrm{s}]$ with $t_{\rm gate}=t_{\rm peak}-t_{<}$ and $t_{<}\in\{0,5,10,20,40,80\}\,t_{M}$.
For the post-merger analysis we gate out $[t_{\rm gate}-1\,\mathrm{s},\,t_{\rm gate}]$ with $t_{\rm gate}=t_{\rm peak}+t_{>}$ and $t_{>}\in\{0,2,4,6,8,10\}\,t_{M_{\rm f}}$. Here $t_{M}$ and $t_{M_{\rm f}}$ are mass-based reference timescales. For GW230814 we adopt the following reference values: $t_{\rm peak}=1376089759.824663\,\mathrm{s}$ (at the geocenter); redshifted total mass $M=65.862\,M_\odot$ corresponding to $t_{M}=0.324\,\mathrm{ms}$; and redshifted final mass $M_{\rm f}=62.731\,M_\odot$ corresponding to $t_{M_{\rm f}}=0.309\,\mathrm{ms}$.

All other analysis settings (e.g., priors, analysis frequency band, reference frequency, and segment duration) follow those used in the official LIGO-Virgo-KAGRA parameter estimates \citep{2025arXiv250818081T}. For the full IMR analyses, the prior on $t_c$ is uniform over a $\pm 0.1\,\mathrm{s}$ window around the trigger time; for the joint pre- and post-truncation analyses, we use a Gaussian prior on $t_c$ centered on the trigger time with standard deviation $0.01\,\mathrm{s}$. To assess modeling systematics we consider four waveform models, \textsc{IMRPhenomXPHM} (XPHM, \citep{2021PhRvD.103j4056P, 2025PhRvD.111j4019C}), \textsc{IMRPhenomXO4a} (XO4a, \citep{2021PhRvD.104l4027H, 2024PhRvD.109f3012T}), \textsc{SEOBNRv5PHM} (v5PHM, \citep{2023PhRvD.108l4038V, 2023PhRvD.108l4037R, 2023PhRvD.108l4036K, 2023PhRvD.108l4035P}), and \textsc{NRSur7dq4} (NRSur, \citep{2019PhRvR...1c3015V}), and when using the \textsc{SEOBNRv5PHM} approximant, we set the maximum waveform frequency equal to the sampling frequency.

The post-merger analysis described above uses the post-merger portion of an inspiral-merger-ringdown (IMR) waveform; we refer to this as the pIMR model. In addition, we model the post-merger (or ringdown) data as a superposition of the quasinormal modes (QNMs) of a perturbed Kerr black hole, using the public \textsc{ringdown} package \citep{2021arXiv210705609I}. Specifically, we include the dominant $\ell=m=2$, $n=0$ QNM (i.e., the fundamental $220$ mode) and/or one of the modes in $\ell mn = \{221,210,200,330,320,440\}$ that expected to have largest amplitudes for BBH mergers \citep{2025arXiv250708219T}. In this QNM analysis, the sky location and polarization angle of GW230814 are fixed to $(\alpha,\delta,\psi) = (1.291,\,-0.356,\,3.127)\,\mathrm{rad}$. The ringdown start time is varied as $t_{\rm start}=t_{\rm peak}+t_{>}$, with $t_{>}\in\{0,2,4,\ldots,20\}\,t_{M_{\rm f}}$ in steps of $2\,t_{M_{\rm f}}$. For GW230814, the prior for the final black hole mass $M_{\rm f}$ is uniform in $[42,86]\,M_\odot$ and the mode amplitudes $A_{\ell mn}$ are uniform in $[0,\,6.166]\times 10^{-20}$.

We analyze open strain data for each detector from the Gravitational-Wave Open Science Center (GWOSC). The strain data are high-pass filtered at the low-frequency cutoff $f_{\rm low}$ with a Butterworth filter, then downsampled to the target sampling rate ($f_s$) and symmetrically cropped to remove filter transients. We estimate the one-sided noise power spectral density (PSD) from a longer stretch of data via Welch method with $8$-s segments and median-mean averaging; we then apply inverse-spectrum truncation and interpolate the PSD to the analysis frequency grid. Throughout our analyses, we use standard GW data-analysis tools. In particular, \textsc{PyCBC} \citep{2019PASP..131b4503B} is used to compute the likelihoods for the joint pre- and post-merger analyses and \textsc{ringdown} package is used to evaluate the QNM likelihoods. We employ \textsc{Bilby} \citep{2019ApJS..241...27A} with \textsc{Dynesty} \citep{2020MNRAS.493.3132S} for nested sampling, adopting an evidence tolerance of \(\Delta\log Z=0.1\) and 2000 live points (other sampler settings follow the \textsc{Bilby} defaults). The code and data used to reproduce our results and generate the figures are openly available on Zenodo \citep{zenodo-url}.

\section{Results}
\begin{figure}
    \centering
    \includegraphics[width=0.49\textwidth]{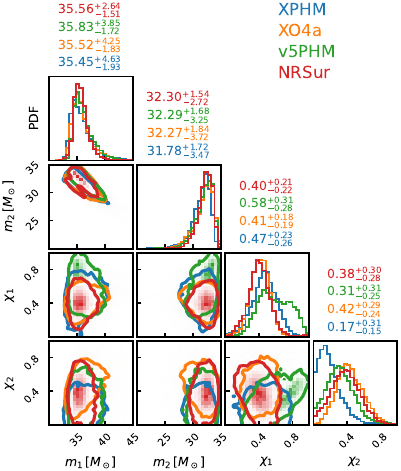}
    \includegraphics[width=0.49\textwidth]{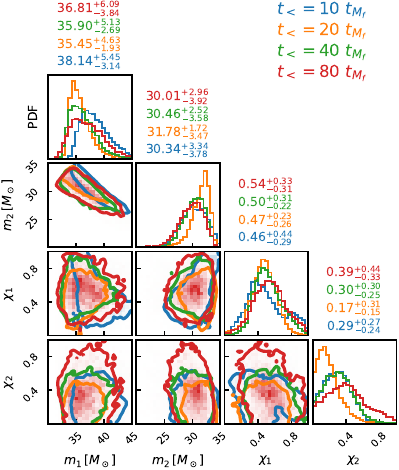}
    \caption{Pre-merger posteriors for the (redshifted) component masses and spins of GW230814. Contours enclose $90\%$ credible regions. \emph{Left}: comparison across waveform models at a fixed truncation $t_{<}=20\,t_{M}$. \emph{Right}: dependence on the pre-truncation $t_{<}$ for XPHM.}
    \label{fig:component-mass-spin}
\end{figure}
Using the pre-merger analyses, we first constrain the component masses and spins for GW230814. We present results for a fiducial pre-truncation choice $t_{<}=20\,t_{M}$ and compare multiple waveform models. As shown in the left panel of Fig.~\ref{fig:component-mass-spin}, the inferred component masses are mutually consistent across different waveforms within posterior uncertainties. The component spins differ slightly; in particular, XPHM shows a mild tendency toward a smaller secondary spin, but the shift remains within statistical uncertainties. We next examine how the choice of pre-truncation $t_{<}$ affects the posteriors, using the XPHM waveform (see the right panel of Fig.~\ref{fig:component-mass-spin}). For $t_{<}=10,20,40\,t_{M}$, the posteriors show no strong dependence on the truncation choice; the slightly tighter constraints at $t_{<}=20\,t_{M}$ likely arise from run-to-run fluctuations. For the larger (earlier) truncation, $t_{<}=80\,t_{M}$ ($\approx 26\,{\rm ms}$), the posteriors broaden, as expected from the loss of SNR when excising more signal.

\begin{figure}
    \centering
    \includegraphics[width=0.49\textwidth]{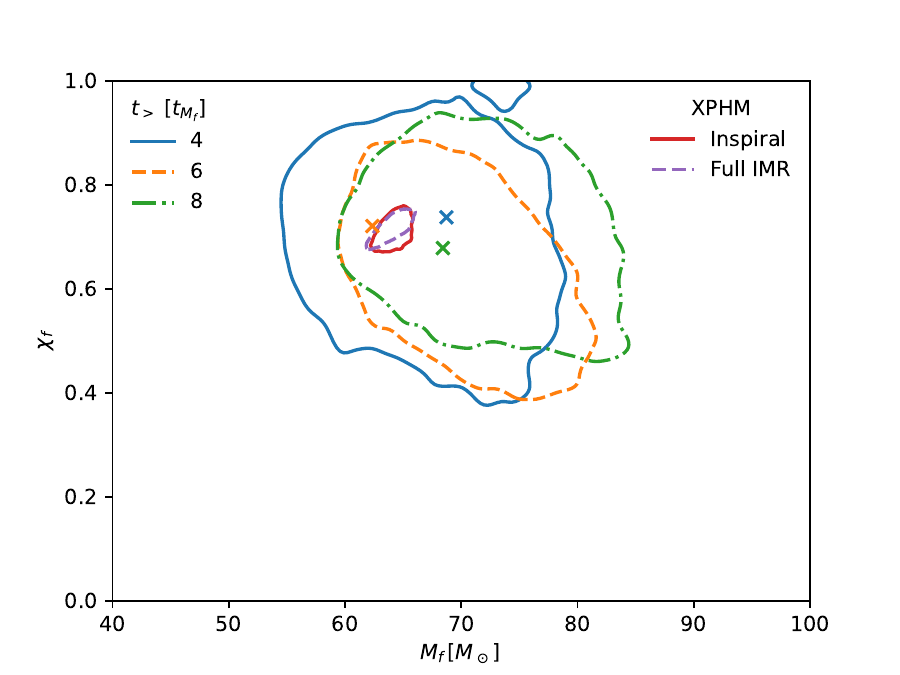}
    \includegraphics[width=0.49\textwidth]{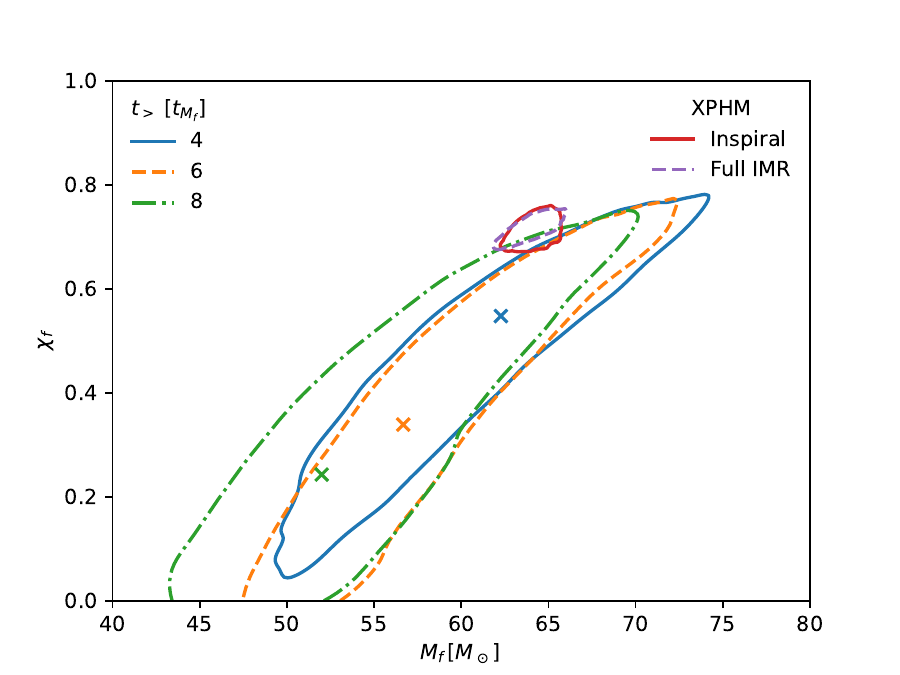}
    \caption{Post-merger posteriors for the (redshifted) remnant mass $M_{\rm f}$ and spin $\chi_{\rm f}$ of GW230814 from pIMR (left) and QNM (right) models. For reference, solid red contours show the reconstructed $M_{\rm f}$ and $\chi_{\rm f}$ from pre-merger analysis using XPHM and $t_{<}=20\,t_{M}$; dotted contours show the predictions from the full IMR analysis. Contours enclose $90\%$ credible regions. Crosses mark the maximum-likelihood points.}
    \label{fig:final-mass-spin}
\end{figure}
Using the post-merger analyses (with pIMR and QNM models), we constrain the remnant mass $M_{\rm f}$ and spin $\chi_{\rm f}$ of GW230814. As shown in the left panel of Fig.~\ref{fig:final-mass-spin}, the resulting posteriors are relatively broad but remain consistent with both the full-IMR and the pre-merger predictions, indicating robustness to moderate choices of $t_{>}=4,6,8\,t_{M_{\rm f}}$. Before presenting the QNM-based estimates of $M_{\rm f}$ and $\chi_{\rm f}$, we first examine whether the data of GW230814 favor including any modes beyond the fundamental 220. We compute the Bayes factor comparing a ringdown model that includes an additional mode (such as the first overtone or a higher harmonic) against the baseline single-mode (220-only) model, as a function of the ringdown start time (as shown in Fig.~\ref{fig:bayes} in the Appendix). In contrast to the case of GW231123 (which showed strong hints of an extra mode in its ringdown \citep{2025arXiv250902047W}), we find that for the GW230814 analyzed here, there is no significant evidence for any mode beyond the fundamental $220$ mode. We therefore adopt the 220-only QNM model to infer $M_{\rm f}$ and $\chi_{\rm f}$. As shown in the right panel of Fig.~\ref{fig:final-mass-spin}, the QNM posteriors exhibit marginal overlap with the IMR/inspiral predictions, owing to a bias toward lower $M_{\rm f}$ and $\chi_{\rm f}$, which becomes more pronounced for later choices of the post-truncation $t_{>}$.

Using the above results, we can now test the area law by directly comparing the total horizon area before and after the merger. For a Kerr black hole of mass $m$ and dimensionless spin $\chi$, the horizon area is given by
\begin{equation}
A = 8\pi m^2 \left( 1 + \sqrt{1 - \chi^2} \right),
\end{equation}
in geometric units ($G=c=1$). From our pre-merger analyses, we obtain posterior samples for the masses ($m_1, m_2$) and spins ($\chi_1, \chi_2$) of the two initial BHs, which we use to compute the total initial horizon area $A_{\rm i} = A(m_1,\chi_1) + A(m_2,\chi_2)$. Similarly, though the post-merger analyses we can also reconstruct samples for the remnant's mass and spin, from which we compute the final horizon area $A_{\rm f} = A(M_{\rm f},\chi_{\rm f})$. Following Ref.~\citep{2022PhRvD.105f4042K}, we compare the \emph{measured} change $A_{\rm f}-A_{\rm i}$ to the \emph{expected} change $A_{\rm f}^*-A_{\rm i}$, where $A_{\rm f}^*$ is the GR prediction obtained by mapping the inspiral-inferred parameters to ($M_{\rm f}$, $\chi_{\rm f}$) via specific waveform models, and define the area ratio $\mathcal{R} = (A_{\rm f}-A_{\rm i})/(A_{\rm f}^*-A_{\rm i})$.

\begin{table}
    \centering
    \caption{Symmetric $68.3\%$ credible intervals for $\mathcal{R}$ and the probabilities for $\mathcal{R}<0$ inferred with pIMR model.}
    \label{tab:pIMR}
    \begin{ruledtabular}
    \begin{tabular}{l|c|ccccc}
    & \diagbox{$t_{>}$}{$t_{<}$} & 5 & 10 & 20 & 40 & 80 \\
    \hline
    \multirow{6}{*}{\rotatebox[origin=c]{90}{XPHM}}
    \rule{0pt}{2.8ex} &0 & $1.20_{-0.43}^{+0.82}$ \ (0.12\%) & $1.20_{-0.43}^{+0.82}$ \ (0.10\%) & $1.28_{-0.43}^{+0.84}$ \ (0.04\%) & $1.31_{-0.42}^{+0.82}$ \ (0.03\%) & $1.29_{-0.43}^{+0.80}$ \ (0.05\%) \\
    &2 & $1.30_{-0.40}^{+0.83}$ \ (69.1 ppm) & $1.30_{-0.39}^{+0.84}$ \ (12.6 ppm) & $1.39_{-0.40}^{+0.86}$ \ (12.9 ppm) & $1.42_{-0.39}^{+0.83}$ \ (35.7 ppm) & $1.40_{-0.40}^{+0.82}$ \ (48.6 ppm) \\
    &4 & $0.93_{-0.40}^{+0.74}$ \ (0.57\%) & $0.93_{-0.40}^{+0.74}$ \ (0.50\%) & $1.01_{-0.40}^{+0.76}$ \ (0.24\%) & $1.04_{-0.40}^{+0.74}$ \ (0.20\%) & $1.03_{-0.40}^{+0.73}$ \ (0.27\%) \\
    &6 & $1.31_{-0.42}^{+0.77}$ \ (0.01\%) & $1.31_{-0.42}^{+0.78}$ \ (20.3 ppm) & $1.40_{-0.42}^{+0.79}$ \ (19.7 ppm) & $1.42_{-0.42}^{+0.77}$ \ (48.9 ppm) & $1.40_{-0.42}^{+0.76}$ \ (61.4 ppm) \\
    &8 & $1.40_{-0.49}^{+0.80}$ \ (0.05\%) & $1.40_{-0.49}^{+0.81}$ \ (0.05\%) & $1.49_{-0.50}^{+0.83}$ \ (0.02\%) & $1.51_{-0.49}^{+0.80}$ \ (0.02\%) & $1.49_{-0.49}^{+0.79}$ \ (0.02\%) \\
    &10 & $1.67_{-0.61}^{+0.79}$ \ (0.03\%) & $1.67_{-0.62}^{+0.79}$ \ (0.03\%) & $1.77_{-0.63}^{+0.81}$ \ (78.9 ppm) & $1.78_{-0.61}^{+0.79}$ \ (82.0 ppm) & $1.75_{-0.60}^{+0.78}$ \ (0.01\%) \\
    \hline
    \multirow{6}{*}{\rotatebox[origin=c]{90}{XO4a}}
    \rule{0pt}{2.8ex} &0 & $1.34_{-0.55}^{+0.84}$ \ (0.03\%) & $1.32_{-0.58}^{+0.87}$ \ (0.08\%) & $1.32_{-0.57}^{+0.86}$ \ (0.07\%) & $1.41_{-0.53}^{+0.80}$ \ (0.01\%) & $1.41_{-0.52}^{+0.77}$ \ (0.07\%) \\
    &2 & $0.76_{-0.29}^{+0.50}$ \ (0.02\%) & $0.72_{-0.30}^{+0.52}$ \ (0.14\%) & $0.73_{-0.30}^{+0.52}$ \ (0.09\%) & $0.86_{-0.29}^{+0.48}$ \ (0.01\%) & $0.89_{-0.31}^{+0.47}$ \ (0.23\%) \\
    &4 & $1.02_{-0.39}^{+0.74}$ \ (0.17\%) & $1.00_{-0.41}^{+0.77}$ \ (0.26\%) & $1.00_{-0.40}^{+0.76}$ \ (0.24\%) & $1.11_{-0.38}^{+0.70}$ \ (0.10\%) & $1.13_{-0.39}^{+0.67}$ \ (0.19\%) \\
    &6 & $1.18_{-0.52}^{+0.77}$ \ (0.10\%) & $1.16_{-0.55}^{+0.81}$ \ (0.22\%) & $1.16_{-0.54}^{+0.80}$ \ (0.18\%) & $1.26_{-0.51}^{+0.74}$ \ (0.05\%) & $1.27_{-0.50}^{+0.71}$ \ (0.16\%) \\
    &8 & $0.99_{-0.35}^{+0.52}$ \ (0.01\%) & $0.96_{-0.37}^{+0.54}$ \ (0.04\%) & $0.96_{-0.36}^{+0.53}$ \ (0.03\%) & $1.07_{-0.34}^{+0.50}$ \ (65.1 ppm) & $1.09_{-0.35}^{+0.49}$ \ (0.09\%) \\
    &10 & $1.05_{-0.39}^{+0.61}$ \ (0.07\%) & $1.02_{-0.41}^{+0.64}$ \ (0.14\%) & $1.02_{-0.40}^{+0.63}$ \ (0.12\%) & $1.13_{-0.38}^{+0.59}$ \ (0.04\%) & $1.15_{-0.39}^{+0.56}$ \ (0.13\%) \\
    \end{tabular}
    \end{ruledtabular}
    \footnotetext{\footnotesize \textbf{Note.} \textit{ppm} = parts per million.
    }
\end{table}

\begin{figure}
    \centering
    \includegraphics[width=0.98\textwidth]{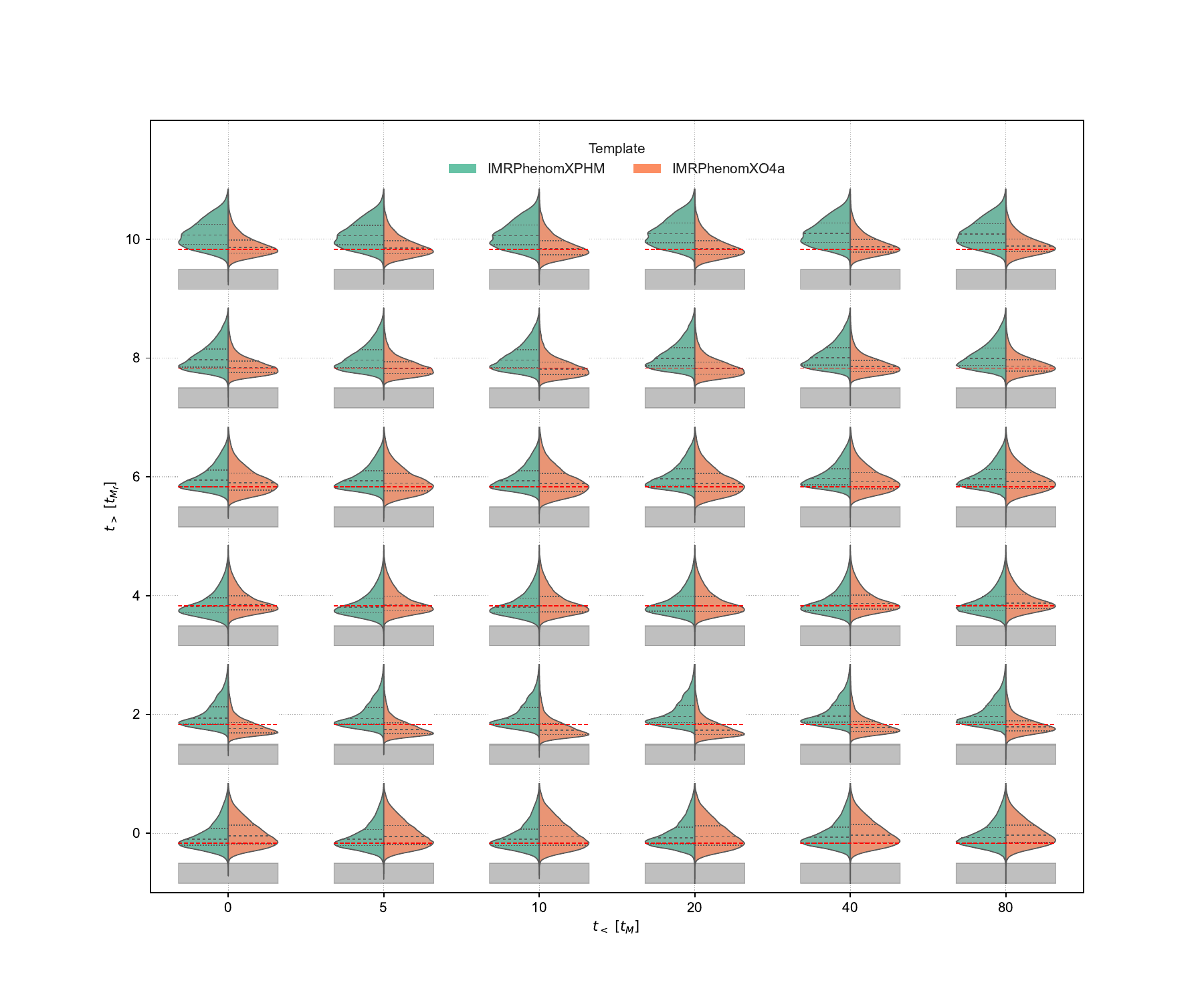}
    \caption{Violin plots of the ratio $\mathcal{R}$ of the measured to the expected change in the black-hole horizon area for multiple combinations of $t_{<}$ and $t_{>}$, obtained with the pIMR analysis. The shaded gray regions indicate violation of the area law ($\mathcal{R}<0$); the red dashed lines ($\mathcal{R}=1$) mark the GR prediction.}
    \label{fig:violin-pIMR}
\end{figure}

\begin{figure}
    \centering
    \includegraphics[width=0.7\textwidth]{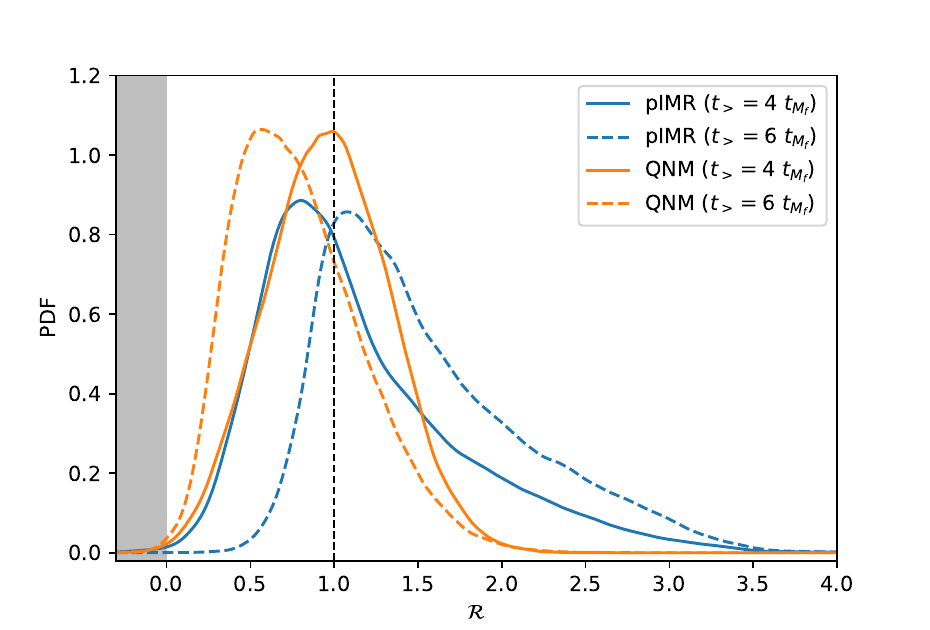}
    \caption{Probability distributions of the ratio $\mathcal{R}$ of the measured to the expected change in the black-hole horizon area for $t_{>}=4,6\,t_{M_{\rm f}}$ (both pIMR and QNM) with $t_{<}=20\,t_{M}$. The distributions lie almost entirely to the right of $\mathcal{R}=0$ (shaded gray), indicating that for these configurations the final horizon area exceeds the initial total area with high probability. The vertical line at $\mathcal{R}=1$ marks the GR prediction.}
    \label{fig:darea-ratio}
\end{figure}
Our pIMR results are summarized in Tab.~\ref{tab:pIMR} and Fig.~\ref{fig:violin-pIMR} (QNM results are presented in the Appendix). For all combinations of $(t_{<},\,t_{>})$ we find $\Delta A \equiv A_{\rm f}-A_{\rm i}>0$ with overwhelming probability. The ratio $\mathcal{R}$ is noticeably more sensitive to the post-truncation $t_{>}$ than to $t_{<}$. Both XPHM and XO4a generally yield $\mathcal{R} \gtrsim 1$, with $\mathcal{R}$ closest to 1 at $t_{>}\approx 4\,t_{M_{\rm f}}$; a notable exception is XO4a at $t_{>}=2\,t_{M_{\rm f}}$, which tends to underestimate $\mathcal{R}$. For the QNM analyses, we observe a systematic evolution with $t_{>}$: as $t_{>}$ increases from $0$ to $10\,t_{M_{\rm f}}$, $\mathcal{R}$ transitions from a mild overestimate to an underestimate ($\mathcal{R}<1$), with $t_{>}\approx 2,4\,t_{M_{\rm f}}$ providing the best agreement with GR. For both pIMR and QNM analyses, choices $t_{>}=4$ and $6\,t_{M_{\rm f}}$ are preferred, as they give $\mathcal{R}$ closest to the GR expectation $\mathcal{R} \simeq 1$. Zoomed-in comparisons at $t_{>}=4,6\,t_{M_{\rm f}}$ for both pIMR and QNM models are shown in Fig.~\ref{fig:darea-ratio}. From Tab.~\ref{tab:pIMR}, focusing on the preferred choices $t_{>}=4$ and $6\,t_{M_{\rm f}}$, the probability of a decrease in total area spans $P(\mathcal{R}<0)\approx 20\,{\rm ppm}$ to $0.57\%$ for GW230814. Thus the area increase is confirmed at $\sim 2.5-4.1\sigma$, strengthening the observational support for the area theorem beyond earlier tests (e.g., GW150914 at $\sim 1.7-2.6\sigma$ \citep{2024PhRvD.110d4018C}).

\section{Summary}
We have tested Hawking's area theorem with the high-SNR event GW230814 by comparing the total horizon area before merger with that of the remnant, using two independent post-merger frameworks: a post-merger IMR (pIMR) model constructed by gating out the inspiral, and a Kerr-QNM ringdown analysis. 
For preferred gate placements, we find $P(\mathcal{R}<0)$ between $\sim 20~\mathrm{ppm}$ and $0.57\%$. This corresponds to a $\sim 2.5-4.1\sigma$ confirmation of Hawking's area theorem, improving upon earlier observational tests. The results are robust to reasonable variations in gate placement, waveform model, and sky location uncertainty.

Our findings have several implications. First, they offer a direct and confidential confirmation of a fundamental aspect of black hole physics. Hawking's area theorem, often likened to the second law of thermodynamics for black holes, is a pillar of classical general relativity. By verifying the area law with such high confidence in real astrophysical mergers, we have reinforced the validity of general relativity in the dynamical strong-field regime. Second, these results demonstrate the increasing power of gravitational-wave observations to test subtle predictions of gravity. Just a few years ago, area-law tests were limited by the moderate confidence level \citep{2021PhRvL.127a1103I, 2022PhRvD.105f4042K, 2024PhRvD.110d4018C}; now, thanks to the improved detectors and the louder GW signals \citep{2025arXiv250818079T, 2025arXiv250818080T, 2025arXiv250818082T}, we can confirm the theorem at a significance exceeding (at least) $\sim 3\sigma$.

Finally, our work paves the way for future explorations. As more data from O4 and upcoming runs (O5 and beyond) become available, we expect to gather a larger sample of high-SNR BBH events, which will further refine the statistics of area-law tests. With a population of events, one could also investigate whether any outlier mergers show anomalous behavior (e.g., a failure of the area law) which could hint at new physics such as exotic compact objects or beyond-GR effects. So far, our observations are fully consistent with the predictions of classical GR. Additionally, next-generation GW detectors (such as Einstein Telescope \citep{2025arXiv250312263A} and Cosmic Explorer \citep{2019BAAS...51g..35R}) will detect BBH mergers with dramatically higher SNR, allowing even more precise tests of the area theorem and other strong-field GR principles. Such observations could potentially detect tiny deviations if any exist, for example, due to quantum gravitational effects that might slightly violate the area law.

\begin{acknowledgments}
This work is supported in part by National Natural Science Foundation of China under grants No. 12233011 and No. 12303056, and by ``the Fundamental Research Funds for the Central Universities" at Dalian University of Technology.
This research has made use of data or software obtained from the Gravitational Wave Open Science Center, a service of LIGO Laboratory, the LIGO Scientific Collaboration, the Virgo Collaboration, and KAGRA \citep{2023ApJS..267...29A}. LIGO Laboratory and Advanced LIGO are funded by the United States National Science Foundation (NSF) as well as the Science and Technology Facilities Council (STFC) of the United Kingdom, the Max-Planck-Society (MPS), and the State of Niedersachsen/Germany for support of the construction of Advanced LIGO and construction and operation of the GEO600 detector. Additional support for Advanced LIGO was provided by the Australian Research Council. Virgo is funded, through the European Gravitational Observatory (EGO), by the French Centre National de Recherche Scientifique (CNRS), the Italian Istituto Nazionale di Fisica Nucleare (INFN) and the Dutch Nikhef, with contributions by institutions from Belgium, Germany, Greece, Hungary, Ireland, Japan, Monaco, Poland, Portugal, Spain. KAGRA is supported by Ministry of Education, Culture, Sports, Science and Technology (MEXT), Japan Society for the Promotion of Science (JSPS) in Japan; National Research Foundation (NRF) and Ministry of Science and ICT (MSIT) in Korea; Academia Sinica (AS) and National Science and Technology Council (NSTC) in Taiwan of China.
\end{acknowledgments}

\appendix
\section{Results for QNM analysis}
\begin{figure}
    \centering
    \includegraphics[width=0.7\textwidth]{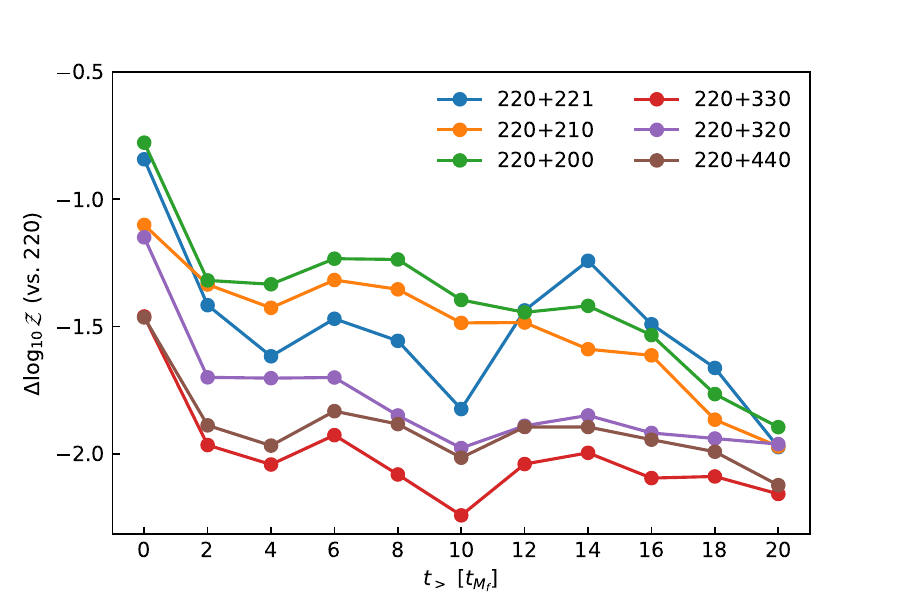}
    \caption{Bayes factors for models including an extra QNM versus the single-mode (220) model, as a function of the chosen $t_{>}$ for GW230814.}
    \label{fig:bayes}
\end{figure}

\begin{table}
    \centering
    \caption{Symmetric $68.3\%$ credible intervals for $\mathcal{R}$ and the probabilities for $\mathcal{R}<0$ inferred with QNM model.}
    \label{tab:QNM}
    \begin{ruledtabular}
    \begin{tabular}{l|c|ccccc}
    & \diagbox{$t_{>}$}{$t_{<}$} & 5 & 10 & 20 & 40 & 80 \\
    \hline
    \multirow{6}{*}{\rotatebox[origin=c]{90}{XPHM}}
    \rule{0pt}{2.8ex} &0 & $1.07_{-0.36}^{+0.39}$ \ (0.06\%) & $1.07_{-0.36}^{+0.38}$ \ (89.7 ppm) & $1.15_{-0.37}^{+0.39}$ \ (95.4 ppm) & $1.18_{-0.36}^{+0.38}$ \ (0.02\%) & $1.16_{-0.36}^{+0.38}$ \ (0.02\%) \\
    &2 & $0.87_{-0.38}^{+0.37}$ \ (0.70\%) & $0.87_{-0.38}^{+0.37}$ \ (0.70\%) & $0.96_{-0.38}^{+0.38}$ \ (0.19\%) & $0.99_{-0.37}^{+0.37}$ \ (0.15\%) & $0.98_{-0.38}^{+0.37}$ \ (0.25\%) \\
    &4 & $0.88_{-0.38}^{+0.37}$ \ (0.63\%) & $0.88_{-0.37}^{+0.37}$ \ (0.61\%) & $0.96_{-0.38}^{+0.38}$ \ (0.18\%) & $0.99_{-0.37}^{+0.37}$ \ (0.13\%) & $0.98_{-0.37}^{+0.37}$ \ (0.22\%) \\
    &6 & $0.67_{-0.33}^{+0.43}$ \ (1.19\%) & $0.66_{-0.33}^{+0.43}$ \ (1.12\%) & $0.74_{-0.33}^{+0.43}$ \ (0.24\%) & $0.78_{-0.32}^{+0.42}$ \ (0.20\%) & $0.77_{-0.33}^{+0.42}$ \ (0.38\%) \\
    &8 & $0.43_{-0.34}^{+0.45}$ \ (9.89\%) & $0.43_{-0.35}^{+0.45}$ \ (10.74\%) & $0.49_{-0.34}^{+0.46}$ \ (6.09\%) & $0.54_{-0.34}^{+0.45}$ \ (4.04\%) & $0.54_{-0.35}^{+0.44}$ \ (4.80\%) \\
    &10 & $0.43_{-0.35}^{+0.47}$ \ (9.79\%) & $0.43_{-0.36}^{+0.47}$ \ (10.70\%) & $0.50_{-0.35}^{+0.48}$ \ (5.97\%) & $0.55_{-0.34}^{+0.47}$ \ (3.98\%) & $0.54_{-0.35}^{+0.46}$ \ (4.76\%) \\
    \hline
    \multirow{6}{*}{\rotatebox[origin=c]{90}{XO4a}}
    \rule{0pt}{2.8ex} &0 & $1.16_{-0.35}^{+0.37}$ \ (2.29 ppm) & $1.14_{-0.37}^{+0.39}$ \ (32.3 ppm) & $1.14_{-0.37}^{+0.39}$ \ (45.4 ppm) & $1.24_{-0.34}^{+0.36}$ \ (10.6 ppm) & $1.25_{-0.35}^{+0.36}$ \ (0.04\%) \\
    &2 & $0.97_{-0.37}^{+0.36}$ \ (0.09\%) & $0.94_{-0.39}^{+0.38}$ \ (0.27\%) & $0.94_{-0.38}^{+0.38}$ \ (0.21\%) & $1.06_{-0.36}^{+0.35}$ \ (0.03\%) & $1.07_{-0.37}^{+0.36}$ \ (0.18\%) \\
    &4 & $0.98_{-0.36}^{+0.36}$ \ (0.08\%) & $0.95_{-0.38}^{+0.38}$ \ (0.23\%) & $0.95_{-0.37}^{+0.37}$ \ (0.18\%) & $1.06_{-0.35}^{+0.35}$ \ (0.03\%) & $1.08_{-0.36}^{+0.35}$ \ (0.17\%) \\
    &6 & $0.76_{-0.32}^{+0.42}$ \ (0.09\%) & $0.72_{-0.33}^{+0.44}$ \ (0.35\%) & $0.73_{-0.33}^{+0.43}$ \ (0.24\%) & $0.86_{-0.32}^{+0.40}$ \ (0.03\%) & $0.89_{-0.33}^{+0.40}$ \ (0.31\%) \\
    &8 & $0.53_{-0.34}^{+0.44}$ \ (4.24\%) & $0.48_{-0.35}^{+0.46}$ \ (7.20\%) & $0.49_{-0.34}^{+0.46}$ \ (6.48\%) & $0.63_{-0.34}^{+0.43}$ \ (1.85\%) & $0.67_{-0.35}^{+0.42}$ \ (2.17\%) \\
    &10 & $0.53_{-0.34}^{+0.46}$ \ (4.16\%) & $0.48_{-0.35}^{+0.48}$ \ (7.11\%) & $0.49_{-0.35}^{+0.48}$ \ (6.36\%) & $0.64_{-0.34}^{+0.44}$ \ (1.81\%) & $0.68_{-0.35}^{+0.43}$ \ (2.16\%) \\
    \end{tabular}
    \end{ruledtabular}
    \footnotetext{\footnotesize \textbf{Note.} \textit{ppm} = parts per million.
    }
\end{table}

\begin{figure}
    \centering
    \includegraphics[width=0.98\textwidth]{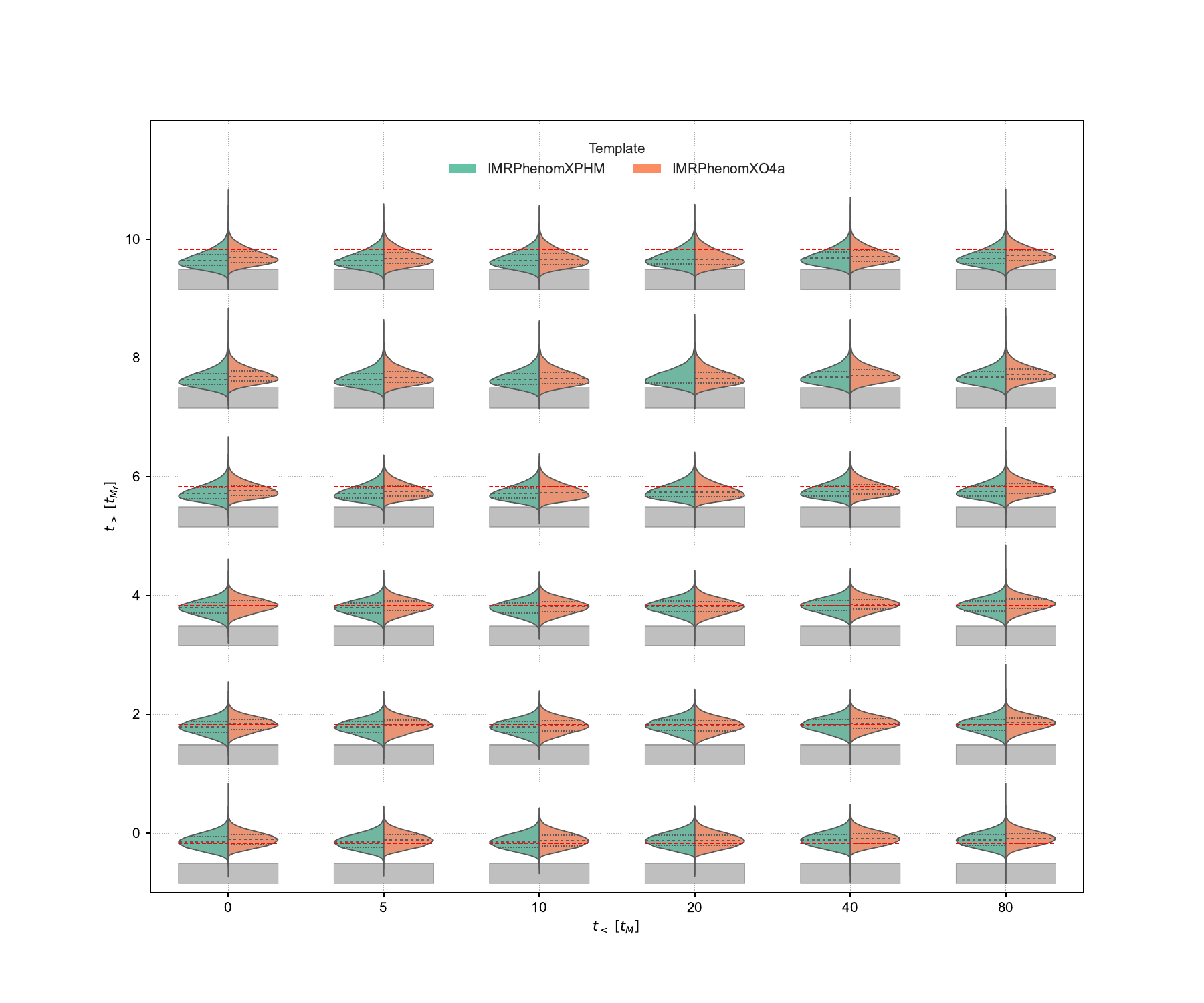}
    \caption{Same as Fig.~\ref{fig:violin-pIMR}, but for the QNM analysis.}
    \label{fig:violin-QNM}
\end{figure}

\bibliographystyle{apsrev4-1}
\bibliography{bibtex.bib}

\end{document}

%% file: aas_macros.tex
\def\apj{ApJ~}                 
\def\apjl{ApJL~}                
\def\apjs{ApJS~}               

\def\jcap{J Cosmology Astropart Phys~}
\def\mnras{MNRAS~}             
\def\prd{Phys~Rev~D~}        
\def\prl{Phys~Rev~Lett~}    
\def\pasp{PASP~}               
